\documentclass{article}
\usepackage[utf8]{inputenc}

\usepackage[english]{babel} 

\usepackage{authblk}

\usepackage{graphicx}
\usepackage{dirtytalk}
\usepackage{longtable}
\usepackage{physics}
\usepackage{amsmath,amssymb}
\usepackage{tikz}
\usetikzlibrary{quantikz}
\usepackage{caption}
\usepackage{hyperref}
\usepackage{colortbl}

\usepackage{biblatex}

\usepackage{epstopdf} 

\addto\captionsenglish{%
}

\title{Comparison between the Iterative Local Search and Exhaustive Search methods applied to QAOA in Max-Cut and Ising Spin Model problems}
\author[1]{M.C.S. Brian Garc\'ia Sarmina}

\affil[1]{Universidad Aut\'onoma Metropolitana, Unidad Azcapotzalco, Ciudad de M\'exico, M\'exico. e-mail:
brian.garsar.6@gmail.com}

\begin{document}

\maketitle

\begin{abstract}
A comparison is made between Exhaustive Search (ES) and Iterative Local Search (ILS). Such comparison was made using the Quantum Approximation Optimization Algorithm (QAOA). QAOA has been extensively researched due to its this potential to be implemented in actual quantum hardware, and its promising future in optimization problems and quantum machine learning. ES and ILS approaches were simulated to determine the pros and cons of these techniques for QAOA in local (classic computer) and real simulations (IBM quantum computer). These classic approaches were used in QAOA to approximate the optimal expected value in Max-Cut and Ising Spin Model (ISM) problems, both of these flavors have three simulated configurations called: linear, cyclic and complete (or full).
\end{abstract}


\section{Introduction}

Nowadays, quantum computing algorithms applied to all kinds of Optimization or Big data problems are of great importance. The actual quantum hardware is considered to be low or intermediate capacity, where most of the devices are not fault-tolerant, which is one of the reasons that quantum algorithms are not regarded as good candidates to be used in actual applications. \cite{1} \cite{2} \cite{10}

For the case of the Quantum Approximation Optimization Algorithm the idea is different, this algorithm is considered a classic-quantum hybrid where some stages of the algorithm are quantum and some are classic, but the difference and the interest of this approach from other quantum algorithms is that QAOA is a low-depth algorithm, in other words, this algorithm does not need too much qubits and quantum gates to operate (only the necessary qubits to represent the problem), and also the QAOA has great fault-tolerant characteristics and due to the low-depth the decoherence factor does not affect considerably the implementation of the algorithm. \cite{3} \cite{4} \cite{7}

Combinatorial optimization problems can be seen in several areas from hardware verification to artificial intelligence, some examples of these problems include the Knapsack problem, Traveling Salesman, Vehicle Routing, etc. These problems can fall into one of these categories: Max-Sat, Max-Cut, and Max-Clique, and as a general approximation we can represent the maximization problem in general. \cite{1} \cite{4} \cite{11}

Let:

\begin{itemize}
    \item There are $n$ binary decision variables $z_{i}$.
    \item $m$ binary functions (called clauses) of those variables contained in $C_{\alpha}(z)$.
    \item And $z_{i} \in \left \{  0, 1 \right \} \forall i \in \left \{1, 2, … , n\right \}$
\end{itemize}

\begin{equation}
    maximize: \sum_{\alpha=1}^{m}C_{\alpha}(z)
\end{equation}

For the case of the \textbf{ISM} (or Ising Energy Model) it could be seen as an \say{extension} of the Max-Cut kind of problem where we also consider an extra part for the cost function. The classic Max-Cut problems only consider the connections between a set of nodes in a graph, but in ISM the presence of an external element known as the \textit{external magnetic field} is also taken into account (these magnetic fields operate individually for every particle or node in the system).

\subsection{Why QAOA ?}

QAOA is considered a near-term quantum algorithm, this stands for an algorithm which can run in a small quantum computer (like IBM Q computers), this algorithm can solve useful problems, etc.

QAOA is a low-depth algorithm, in other words this algorithm does not need numerous quantum gates to be applicable, and this characteristics means that the algorithm shouldn't need too much coherence, and finally some results show that QAOA is fairly robust against errors. \cite{3} \cite{4}

The number of qubits needed to develop the algorithm, in theory, are the number of qubits to store the problem information.

\subsection{Interpretation of C(z) in a QAOA Problem}

As a general example for a problem that can be solved using QAOA is a problem with some set of variables which create a large space of combinations, where one of those combinations is our best result, is the combination which optimizes some criteria. This \say{criteria} is related to the Cost Function denoted by $C(z)$, where $z$ contains all the set of variables to $n$ be optimized.

\begin{equation}
    z \equiv \left \{ z_{1}, z_{2}, ... , z_{n} \right \} ; z_{j} \left \{ -1, +1 \right \}
\end{equation}

Now, the problem is: how we translate the $C(z)$ cost function into a valid quantum problem, this \say{translation} of the problem consists in creating an operator $C$ which will return a value represented by the cost function. This operator $C$ can be seen as the diagonal matrix of $C(z)$ where each value in the diagonal is the possible values of $C(z)$.

\begin{equation}
    C(z)\rightarrow C \ket{z} = C\ket{z_{1}z_{2}...z_{n}} = C(z)\ket{z_{1}z_{2}...z_{n}}
\end{equation}

Now, the representation for the $z_{j}$ variables can be expressed in a \say{quantum computing} way using the Pauli $Z$ operator, this operator follows the property that every value of $z$ should be ${+1,-1}$, the representation of the $Z$ operator is as follows.

\begin{equation*}
    Z = \begin{pmatrix}
            1 & 0\\ 
            0 & -1
        \end{pmatrix}
    ; I = \begin{pmatrix}
            1 & 0\\ 
            0 & 1
        \end{pmatrix}
\end{equation*}

With this representation of $Z$ in mind, now it can be established the new representation for the cost function using the $Z$ operator, the identity operator $I$ and tensor products between them.

\begin{equation}
    C(Z) =  Z_{1} \otimes Z_{2} \otimes ... \otimes Z_{n} \otimes I_{n}
\end{equation}

Depending on the problem, the sequence and how the tensor products of $Z$ and $I$ are applied will be different.

\begin{equation*}
    C(Z) = \begin{bmatrix}
1 & 0\\ 
0 & -1
\end{bmatrix}_{1} \otimes
\begin{bmatrix}
1 & 0\\ 
0 & -1
\end{bmatrix}_{2} \otimes ...
\end{equation*}

\begin{equation*}
    ...\otimes
\begin{bmatrix}
1 & 0\\ 
0 & -1
\end{bmatrix}_{n} \otimes
\begin{bmatrix}
1 & 0\\ 
0 & 1
\end{bmatrix}_{n}
\end{equation*}

And this tensor products of the operators $Z_{n}$ and $I_{n}$ result on the diagonal matrix with the $C(z)$ expected values.

\begin{equation}
    C(Z) = \begin{bmatrix}
 C(Z_{1}) &  0&  0&  0&  0&  0 \\ 
 0 &  C(Z_{2}) &  0&  0&  0&  0 \\ 
 0 &  0&  . &  0&  0&  0 \\ 
 0 &  0&  0&  . &  0&  0 \\ 
 0 &  0&  0&  0&  . &  0 \\ 
 0 &  0&  0&  0&  0& C(Z_{n})  \\ 
\end{bmatrix}
\end{equation}

The matrix representation allows identifying where the $C(Z)$ values appear, where each $C(Z_{i})$ element corresponds to a particular outcome from the classic cost function. After the diagonalization of the cost function, the matrix need to be unitary to be a valid quantum operator \cite{2} \cite{5}

\subsubsection*{Phase Operator $U(C,\gamma)$}

So the next step is to convert $C(Z)$ into a unitary operator, this process can be achieved using the \textbf{Phase Operator} $U(C,\gamma)$. The Phase Operator is responsible to \say{translate} the optimization problem into a valid quantum operator, this operator establishes the connections between particles or nodes (depending on the type of problem analyzed) and the constraint factors (like in ISM where a function related to the external magnetic field is needed), which is defined as:

\begin{equation}
    U(C,\gamma)= e^{-i \gamma C}
\end{equation}

The application of the phase operator also guarantees that the resulting matrix will be diagonal, but more important due to the fact that $U(C, \gamma)$ contains the imaginary element $i$ the resulting matrix elements will also contain the $i$ element. With these two properties of $C(Z)$ being diagonal and having the $e^{i}$ elements guarantees that the resulting matrix will be unitary.

\begin{equation*}
    U(C,\gamma)= \begin{bmatrix}
 e^{i C(Z_{1})  \gamma} &  0&  0&  0&  0&  0\\ 
 0 &  e^{i C(Z_{2}) \gamma} &  0&  0&  0&  0\\ 
 0 &  0&  . &  0&  0&  0\\ 
 0 &  0&  0&  . &  0&  0\\ 
 0 &  0&  0&  0&  . &  0\\ 
 0 &  0&  0&  0&  0&   e^{i C(Z_{n}) \gamma}
\end{bmatrix}
\end{equation*}

In the equation from above, it can be seen that every element from the diagonal matrix form of the $C(Z)$ function is contained in the matrix $U(C,\gamma)$ in the form of phases.

In the case of the matrix of $C(Z)$, a new term is found $\gamma$, this value does not come from the application of the phase gate purely, this $\gamma$ variable is related to the optimization part of the QAOA, and as a brief mention, it will be added another variable named $\beta$, these two variables are the ones that will give us the superposition state which after a measurement should yield with \textbf{high probability} to the state that solves the optimization problem.

\begin{equation}
    U(C,\gamma)\ket{s} = \frac{1}{2}(e^{iC(Z_{1})\gamma}\ket{0}^{\otimes N} + e^{iC(Z_{n})\gamma}\ket{1}^{\otimes N}
    \label{eq:cz_phae_operator}
\end{equation}

\subsubsection*{Mixing Operator $U(B,\beta)$}

After the application of the phase operator, every value of the original cost function is distributed as encoded phases in the $e^{-iC\gamma}$ terms seen on the \textbf{Equation \ref{eq:cz_phae_operator}}, but even if this representation is indeed unitary, when we measure this state every element has the same probability of occurrence, and this equal probability is not one of the thing we want from QAOA, we want QAOA to give us the optimum state with greater probability than the other possible states.

With this idea in mind, the next operator to be applied is the \textbf{Mixing Operator}, this operator has the property of \say{mixing} all the amplitudes of the possibles states (e.g. \textbf{Equation \ref{eq:cz_phae_operator}}) to create constructive and destructive interference, the interference pattern relate higher or lower amplitudes (and hence the probability) of every state.

Let:

\begin{itemize}
    \item $X_{j}$ is the $X$ Pauli operator, and $j=1,2,...,N$.
        \begin{equation*}
            X = \begin{bmatrix}
                    0 && 1 \\
                    1 && 0
                \end{bmatrix}
        \end{equation*}
    \item $B$ is composed of $X$ operators.
    \item $\beta$ is the \say{mixing} free parameter.
\end{itemize}

\begin{equation}
    B = \sum_{j}^{N}X_{j}
\end{equation}

\begin{equation}
    U(B, \beta) = e^{i \beta B}
\end{equation}

The difference between the phase operator and the mixing operator, is that after the application of the exponentiation of the operator $B$ the result is no longer a diagonal matrix operator (due to the structure of the $X$ Pauli), and this led to the amplitude mixing.

Because $B$ is a linear sum of $X$ operator applied for each qubit in the system, the mixing operator $U(B, \beta)$ can be represented as follows:

\begin{equation}
    U(B, \beta) = \coprod_{j}^{N}e^{i\beta X_{j}}
    \label{eq:mo_coprod}
\end{equation}

In the \textbf{Equation \ref{eq:mo_coprod}} the $N$ corresponds to the number of qubits in the system. Each $e^{i\beta X_{j}}$ has the following structure:

\begin{equation}
    e^{i\beta X} = R_{X}(\beta) = \begin{bmatrix}
                        cos(\beta) && i sin(\beta) \\
                        i sin(\beta) && cos(\beta)
                    \end{bmatrix}
\end{equation}

\subsubsection*{Expectation Value $F(\gamma,\beta)$}

In general, the goal of QAOA is to minimize or maximize the expectation value of the cost function. This expectation value can be represented using the following equations.

\begin{equation}
    \ket{\psi_{\gamma \beta}}=U(B,\beta) U(C,\gamma) \ket{s}
    \label{eq:expectation_value}
\end{equation}

\begin{equation}
    F(\gamma,\beta)=\bra{\psi_{\gamma \beta}}C(Z)\ket{\psi_{\gamma \beta}}
\end{equation}

The expectation value is obtained by repeated measurements over the same prepared quantum state, this repeated measurements try to approximate the wave-function of the system. The measurement process is repeated several times to improve the accuracy of the expectation value. 

Sometimes the amplitudes related to the wave-function are said to be the \say{real} amplitude values for the possible state outcomes, and the measured values are called the \say{approximation}, as it was said, with the increase of the number of experiment measurements the accuracy increases but also the length of the algorithm increase and the execution time.

\subsection{Max-Cut}

Max-Cut problems are one of the most interesting and popular problems to implement QAOA on because Max-Cut is a problem that has been widely studied by a lot of researches of different fields in the past decades. Nowadays, the study of the Max-Cut problem has been leaded to the application of quantum algorithms, in specific, QAOA is one of the most promising ones. The general idea of the problem is to create two subsets of nodes from a general set of nodes (graph) where the connections between each subset is the maximum possible. \cite{8} \cite{9}

\begin{figure}[ht]
\centering
\captionsetup{justification=centering}
\includegraphics[width=8cm,height=5cm]{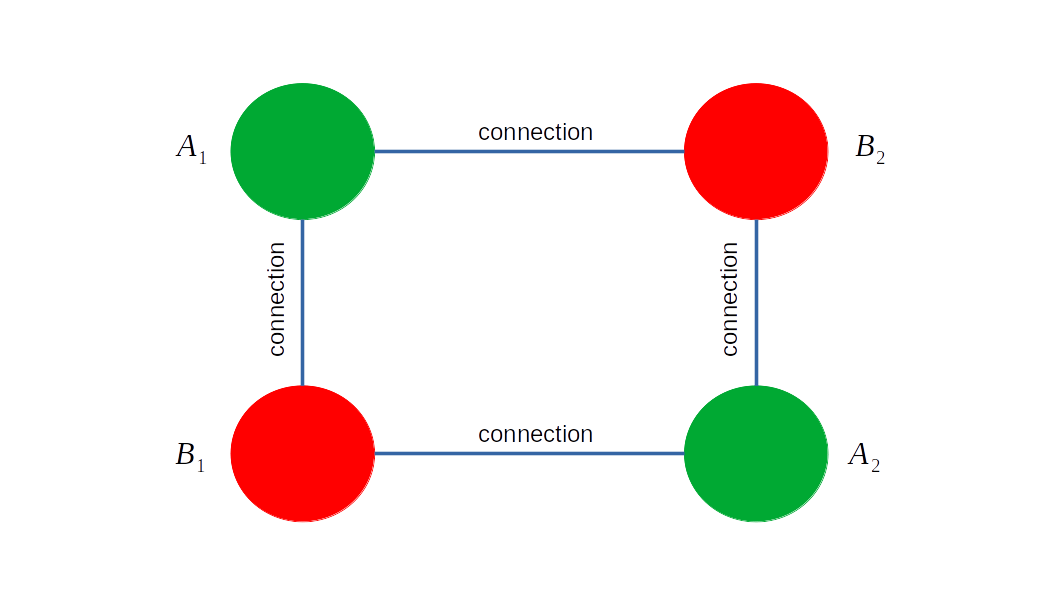}
\caption{Max-Cut problem example with optimal solution.}
\label{fig:MC_PROBLEM}
\end{figure}

The \textbf{Figure \ref{fig:MC_PROBLEM}} is used as an example of how Max-Cut problems are solved, the main graph contains all the nodes $A_{1}$, $A_{2}$ and $B_{1}$, $B_{2}$; each node has one or more connections with the other ones (mostly depending on the configuration of the graph), the idea of how to solve the problem is to separate the nodes in two subsets (the green subset and the red subset) and count how many connections are between the two subsets, the figure from above show the one of the optimal solutions that is having the nodes $A_{1}$ and $A_{2}$ in the subset of green nodes and having the $B_{1}$ and $B_{2}$ in the subset of red nodes (the other optimal configuration is to exchange both nodes of the green subset to the red subset and vice versa). Normally these problems have at least two solutions (an optimal state and the inverse which only changes the \say{labels} between the subsets).

\begin{equation}
    U(H_{p}) = U(C,\gamma)_{MC} = -\sum_{\left \langle i,j \right \rangle} J_{ij} z_{i}z_{j}
    \label{eq:MC_phase_problem}
\end{equation}

The simple form for Max-Cut problem has only one consideration for the description of the problem, that is the connections between nodes, and the \textbf{Equation \ref{eq:MC_phase_problem}} represents these connections between nodes.

\begin{equation*}
    U(C, \gamma) = e^{-i\gamma C} = e^{-i\gamma (\sum_{\left \langle i,j \right \rangle}Z_{i}Z_{j}) }
\end{equation*}

\begin{equation*}
    = e^{-i\gamma (\sum_{\left \langle i,j \right \rangle} Z_{i}Z_{j})}
\end{equation*}

\begin{equation}
    = \prod_{\left \langle i,j \right \rangle}e^{-i\gamma Z_{i}Z_{j}}
    \label{eq:MC_phase_operator}
\end{equation}

The \textbf{Equation \ref{eq:MC_phase_operator}} represents the phase operator, where it is only considered the interaction (connection) between nodes of the graph, every pair of $\left \langle i,j \right \rangle$ elements corresponds to a pair of nodes ($node_{i}$ connects to $node_{j}$) which share a connection between them, this connection does not have any specific direction.

Also, as the example from above, we can replace $z_{i}$ and $z_{j}$ for the $Z$ operator where $Z_{i}$ and $Z_{j}$ can only take values from ${-1,+1}$. Seeing the \textbf{Equation \ref{eq:MC_phase_operator}} it can be established that the neighbors who have the same \say{spin direction} result in a $+1$ for the first summation on the equation, and the result is $-1$ if the nodes have different spin orientation. 

The \say{spin orientation} is related to the way the nodes are separated, in other words if a qubit (node) is on a state $\ket{0}$ (one spin orientation) it can be interpreted as the qubit been on a specific subset and if the qubit is on the state $\ket{1}$ (another spin orientation) it will represent that the qubit is on the opposite subset.

The representation for the mixing operator is the following:

\begin{equation}
    C(U, \beta) =  \prod_{ i } e^{-i \beta X_{i}}
    \label{eq:mixing_operator_MC}
\end{equation}

The \textbf{Equation \ref{eq:mixing_operator_MC}} is understood as the application of the mixing operator (due to the matrix form of the $X_{i}$ quantum gate) which generates the constructive or destructive interference patterns for the states of the system.

\subsubsection{Max-Cut problems}

The specific problems that explored of the class Max-Cut have three configurations, the first configuration corresponds to a $3$ ($P_{1}$, $P_{2}$ and $P_{3}$) nodes with linear configuration.

\begin{figure}[ht]
\centering
\captionsetup{justification=centering}
\includegraphics[width=8cm,height=5cm]{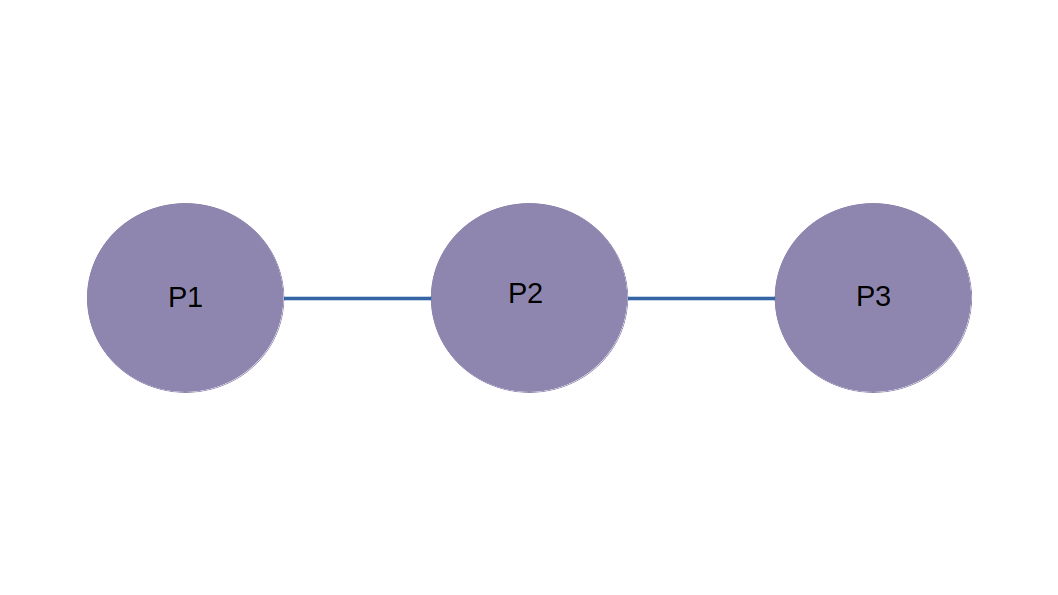}
\caption{Max-Cut $3$ nodes with linear configuration problem.}
\label{fig:MC_3n_LIN_PROBLEM}
\end{figure}

The \textbf{Figure \ref{fig:MC_3n_LIN_PROBLEM}} represents the linear problem, in this case the problem has three nodes with connections $P_{1}$ with $P_{2}$ and $P_{2}$ with $P_{3}$. The optimal solutions for the problem (represented using \textit{BRA-KET} notation for the state that is going to be obtained) are $\ket{010}$ and the inverse $\ket{101}$, these states have the nodes $P_{1}$ and $P_{3}$ in one subset and the node $P_{2}$ on the opposite subset. The evaluation value for the optimal states is $2$.

\begin{figure}[ht]
\centering
\captionsetup{justification=centering}
\includegraphics[width=8cm,height=5cm]{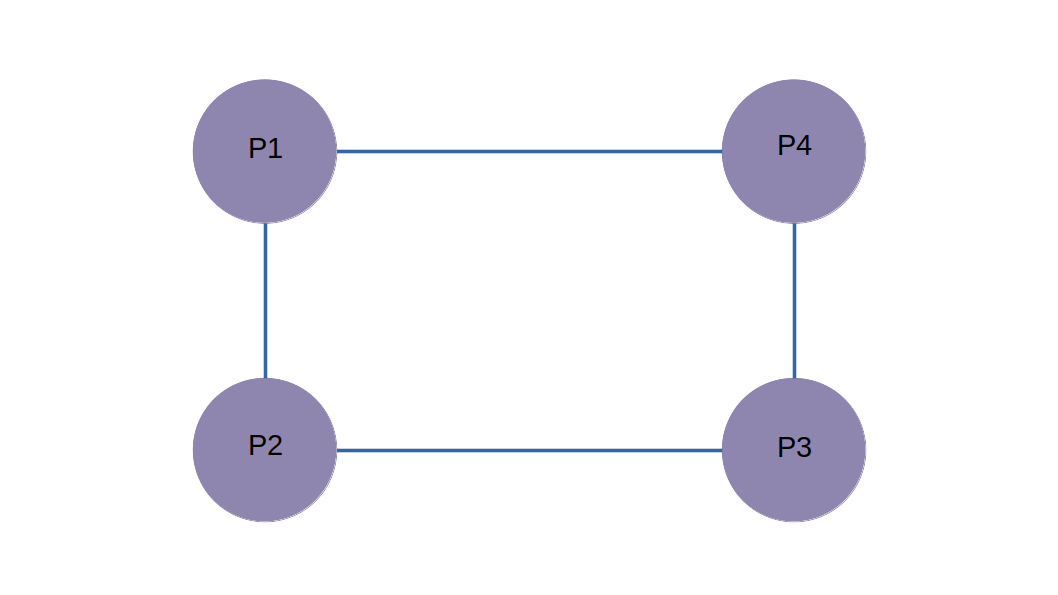}
\caption{Max-Cut $4$ nodes with cyclic configuration problem.}
\label{fig:MC_4n_CYC_PROBLEM}
\end{figure}

The next problem in the \textbf{Figure \ref{fig:MC_4n_CYC_PROBLEM}} which has $4$ nodes in cyclic configuration, due to the configuration this problem has also two optimal state which are the states $\ket{0101}$ and the inverse $\ket{1010}$, the nodes $P_{1}$ and $P_{3}$ are in one subset and the nodes $P_{2}$ and $P_{4}$ belong to the other subset, the expected value evaluation for the optimal states is $4$.

\begin{figure}[ht]
\centering
\captionsetup{justification=centering}
\includegraphics[width=8cm,height=5cm]{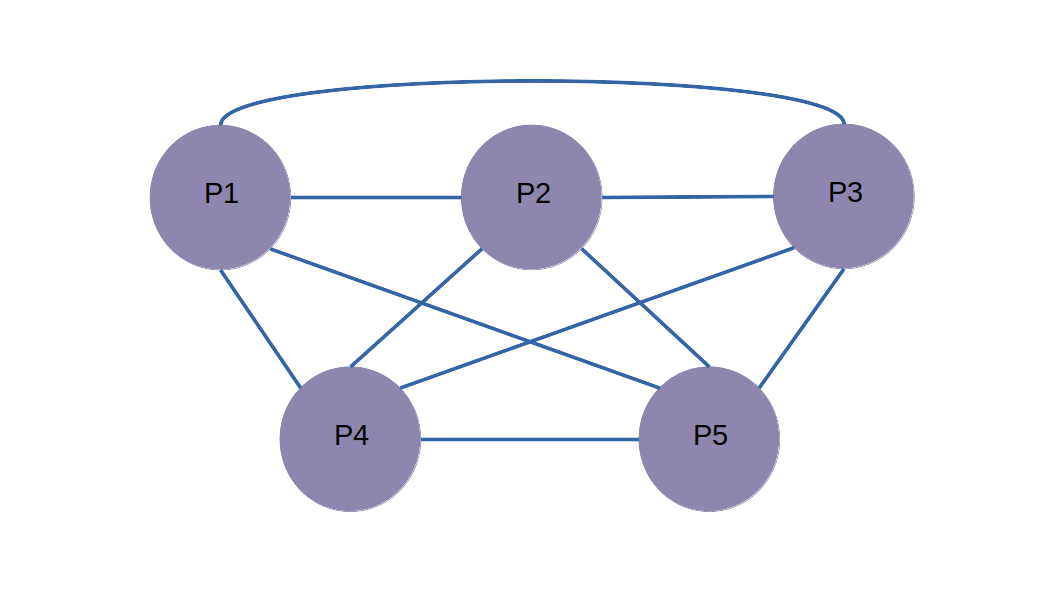}
\caption{Max-Cut $5$ nodes with complete (or full) configuration problem.}
\label{fig:MC_5n_COM_PROBLEM}
\end{figure}

The last instance for the Max-Cut problems is in the \textbf{Figure \ref{fig:MC_5n_COM_PROBLEM}} that represents the $5$ nodes with complete (or full) configuration graph, due to the type of configuration this problem in particular has a lot of optimal state some of them are the states $\ket{00011}$, $\ket{01110}$, $\ket{10110}$, etc. The expected value evaluation for any of the optimal states is $6$.

\subsection{Ising Spin Model}

The Ising Spin Models (ISM) or Ising Energy Models are real examples of problems that can be used to be optimized by QAOA algorithm. These systems are defined by a $d$ dimensional arrangement of $n$ number of spins $z_{i}$, one spin in every \say{lattice cell} $i$, and every particle (with associated spin) is affected by their nearest neighbors and the individual external magnetic field for every particle in the arrangement. The equation called the \textbf{nearest-neighbor Edwards-Anderson Hamiltonian} represents the energy of the system given by the nearest neighbors $\left \langle i,j \right \rangle$, $z_{i}$ and $z_{j}$ are the spins for the couple of particles, $J_{ij}$ is the coupling of $i$ and $j$, and $h_{i}$ is the individual (local) magnetic field for each particle. \cite{6}

\begin{figure}[ht]
\centering
\captionsetup{justification=centering}
\includegraphics[width=8cm,height=5cm]{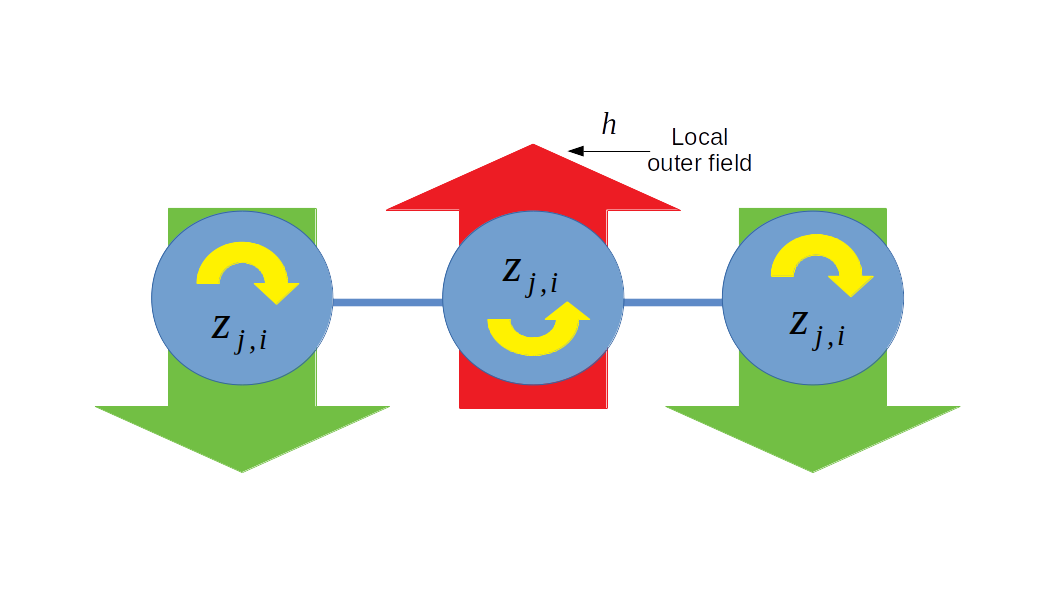}
\caption{Ising Spin Model (ISM) or Ising Energy Model}
\label{fig:ising_spin_model}
\end{figure}

\begin{equation}
    U(H_{p}) = U(C,\gamma)_{ISM} = -\sum_{\left \langle i,j \right \rangle} J_{ij} z_{i}z_{j} - \sum_{i=1}^{n} h_{i} z_{i}
    \label{eq:ising_model}
\end{equation}

The \textbf{Equation \ref{eq:ising_model}} shows the phase operator for the ISM, this phase operator is also known as the \textit{Problem Hamiltonian}, as we said before, the main purpose of the phase operator is to translate the optimization problem to a valid quantum operator.

The optimal solution for these kinds of problems is a certain combination of the spins of the particles which are part of the quantum system, the number of possible combinations of $n$ spins is $2_{n}$ possible combinations, and it can be deduced that the ISM problem increases in complexity very fast as the number of particles in the system increases, and with fairly lager number of $n$ the number of combinations cannot be tracked by classical computers.

The encoded cost function from \textbf{Equation \ref{eq:ising_model}} in terms of a valid matrix operator for quantum computing, is represented in the next formula.

\begin{equation*}
    U(C, \gamma) = e^{-i\gamma C} = e^{-i\gamma (-\sum_{\left \langle i,j \right \rangle}Z_{i}Z_{j}-\sum_{i}h_{i}Z_{i})}
\end{equation*}

\begin{equation*}
    e^{i\gamma (\sum_{\left \langle i,j \right \rangle} Z_{i}Z_{j})} e^{i\gamma(\sum_{i}h_{i}Z_{i})}
\end{equation*}

\begin{equation}
    \prod_{\left \langle i,j \right \rangle}e^{i\gamma Z_{i}Z_{j}} \prod_{i} e^{i\gamma h_{i}Z_{i}}
    \label{eq:ising_coded_function}
\end{equation}

Now, the second part of the preparation for the ISM problem to be compatible with QAOA is to create the mixing operator for the problem, as a general idea all the mixing operators in the most basic form are equal for different problems as it can be seen in the mixing operator for the Max-Cut problem (sign changed).

\begin{equation}
    C(U, \beta) =  \prod_{ i } e^{i \beta X_{i}}
    \label{eq:mixing_operator_ISM}
\end{equation}

\subsubsection{ISM problems}

The problems analyzed for the ISM class are the following, these problems are the direct equivalent problem for the Max-Cut class in the sense that the problems are $3$ particles in linear configuration, $4$ particles in cyclic configuration and $5$ particles in complete configuration.

\begin{figure}[ht]
\centering
\captionsetup{justification=centering}
\includegraphics[width=7.5cm,height=4.5cm]{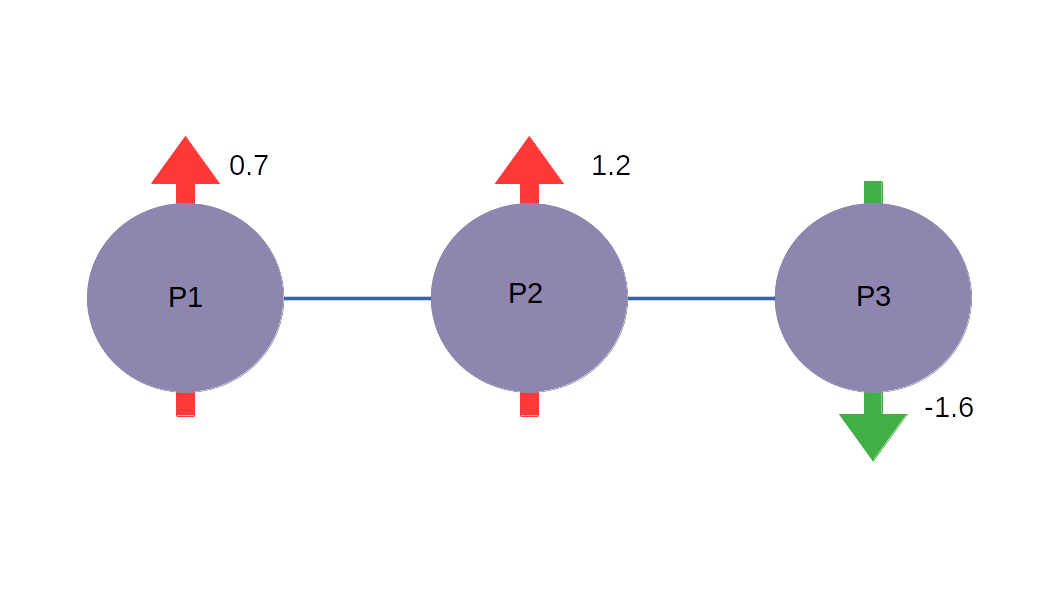}
\caption{ISM $3$ particles with linear configuration problem.}
\label{fig:ISM_3p_LIN_PROBLEM}
\end{figure}

In the \textbf{Figure \ref{fig:ISM_3p_LIN_PROBLEM}} is described the problem for ISM with $3$ particles in linear configuration, the general idea of the problem is relatively similar to the instance for Max-Cut, but in this case the problem acquires a new element that corresponds to the external magnetic field.

\begin{figure}[ht]
\centering
\captionsetup{justification=centering}
\includegraphics[width=8cm,height=5cm]{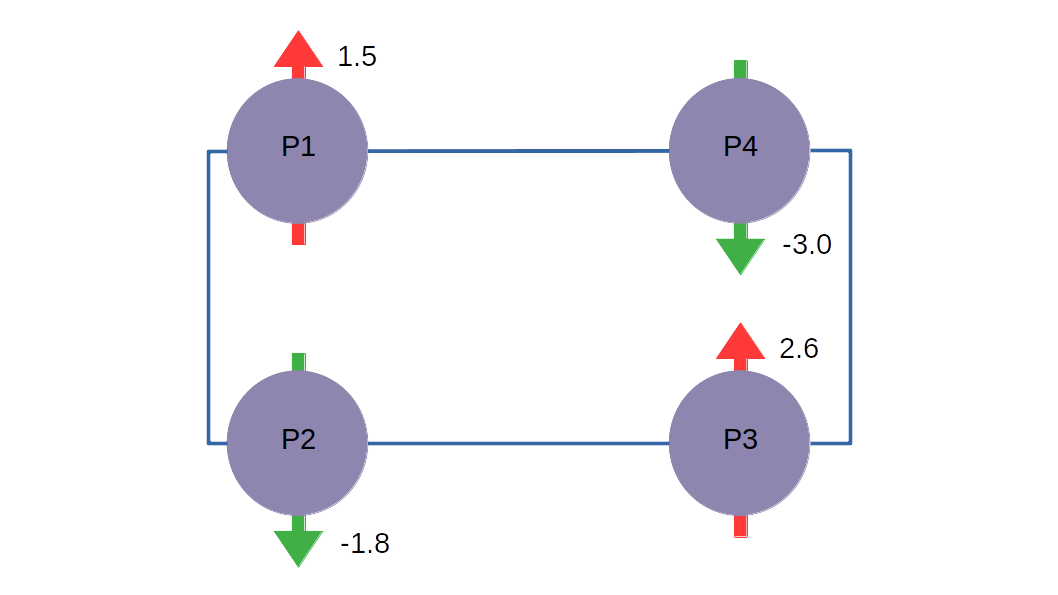}
\caption{ISM $4$ particles with cyclic configuration problem.}
\label{fig:ISM_4p_CYC_PROBLEM}
\end{figure}

The \textbf{Figure \ref{fig:ISM_4p_CYC_PROBLEM}} represents the problem for the $4$ particles case, in this configuration there is a connection between the first and last particle to complete the cycle. 

\begin{figure}[ht]
\centering
\captionsetup{justification=centering}
\includegraphics[width=8cm,height=5cm]{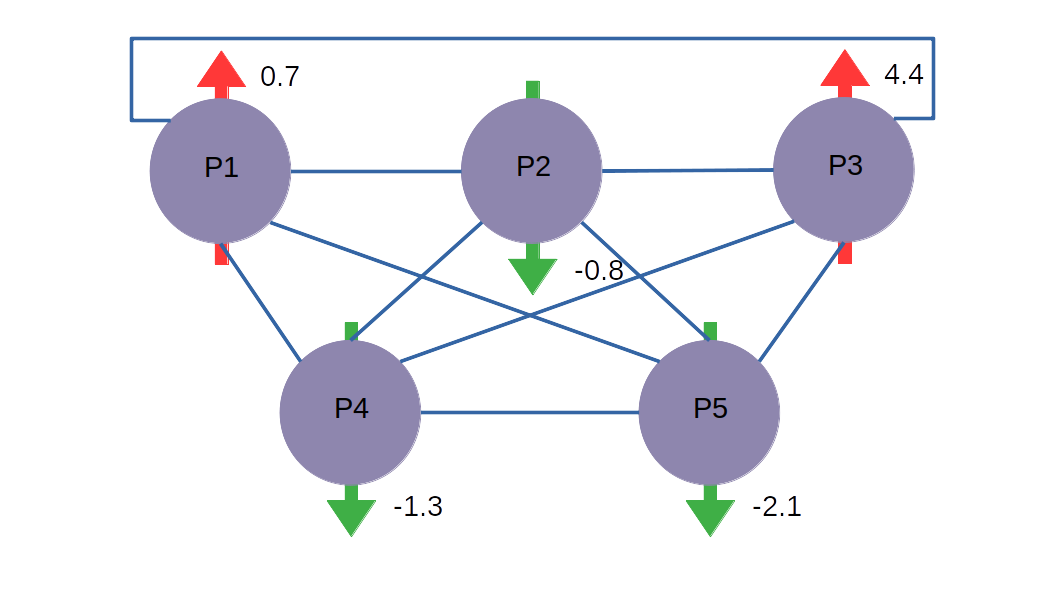}
\caption{ISM $5$ particles with complete (or full) configuration problem.}
\label{fig:ISM_5p_COM_PROBLEM}
\end{figure}

The last problem is shown in the \textbf{Figure \ref{fig:ISM_5p_COM_PROBLEM}}, in this case we have the complete or full configuration of the problem with $5$ particle, the name of the configuration stands for the connections allowed for each node, this configuration establishes that each particle of the system has a direct connection with all the other particles in the system.

\section{General implementation of QAOA}

After obtaining the \textbf{Phase Operator} $U(C,\gamma)$ and the \textbf{Mixing Operator} $U(B,\beta)$, the general preparation of the QAOA is listed below.

\begin{enumerate}
    \item Prepare the initial state: $\ket{\psi}=H^{\otimes N}\ket{0} = \ket{s}$, where $N$ is the number of qubits in the system.
    \item Select the free parameters $\gamma$ and $\beta$ (free or implemented by a function).
    \item Apply the phase operator and mixing operator: $U(B,\beta) U(C,\gamma) \ket{s}$.
    \item Apply the measurement in the computational basis $\ket{z{}'}$.
    \item Check $C(z{}')$
    \item Repeat 1,3,4,5 and compute the expectation value $F(\gamma,\beta)=\bra{\psi_{\gamma \beta}} C(Z) \ket{\psi_{\gamma \beta}}$.
    \item Search for the optimal values for $\gamma$ and $\beta$ for the step 2.
    \item Based on the classic optimization step, establish the best $C(z{}')$.
\end{enumerate}

Finding the values for $\gamma$ and $\beta$ who maximizes or minimizes (depending on the problem) the $F(\gamma, \beta)$ which will lead to the measurement (with greater probability) of the $\ket{z{}'}$ state which is the state that optimizes our classic cost function, these parameters are found with classical methods, unfortunate nowadays there is no known a good quantum algorithm that can perform the finding steps of these variables. The parameters $\gamma$ and $\beta$ span the full range of rotations from $0$ to $2\pi$.

The quantum circuit for the example in the Max-Cut subsection (\textbf{Figure \ref{fig:MC_PROBLEM}} is showed in the next diagram, the circuit represents the $4$ nodes with cyclic configuration Max-Cut.

\begin{figure}[ht]
    \begin{center}
        \begin{tikzpicture}
            \node[scale=0.8] {
                \begin{quantikz}
                    \ket{0}_{1} \slice{(1)}  & \gate{H} \slice{(2)} & \ctrl{1} &\qw & \ctrl{1} & \qw & \qw & \qw & \qw \\
                    \ket{0}_{2} & \gate{H} & \targ{} & \gate{R_{z}(2\gamma)} & \targ{} & \ctrl{1} & \qw & \ctrl{1} & \qw \\
                    \ket{0}_{3} & \gate{H} & \qw & \qw & \qw & \targ{} & \gate{R_{z}(2\gamma)} & \targ{} & \qw \\
                    \ket{0}_{4} & \gate{H} & \qw & \qw & \qw &
                    \qw & \qw & \qw & \qw
                \end{quantikz}
            };
        \end{tikzpicture}
    \end{center}
\end{figure}

\begin{figure}[ht]
    \begin{center}
        \begin{tikzpicture}
            \node[scale=0.75] {
                \begin{quantikz}
                    \qw & \qw & \qw & \ctrl{3} & \qw & \ctrl{3} \slice{(3)} & \gate{R_{x}(\beta)} \slice{(4)} & \qw & \meter{} \\
                    \qw & \qw & \qw & \qw & \qw & \qw & \gate{R_{x}(\beta)} & \qw & \meter{}\\
                    \ctrl{1} & \qw & \ctrl{1} & \qw & \qw & \qw & \gate{R_{x}(\beta)} & \qw & \meter{} \\
                    \targ{} & \gate{R_{z}(2\gamma)} & \targ{}  & \targ{} & \gate{R_{z}(2\gamma)} & \targ{} & \gate{R_{x}(\beta)} & \qw & \meter{}
                \end{quantikz}
            };
        \end{tikzpicture}
    \caption{Quantum circuit for the $4$ nodes with linear configuration Max-Cut problem.}
    \label{fig:MC_QuantumCircuit_4n_CYC}
    \end{center}
\end{figure}
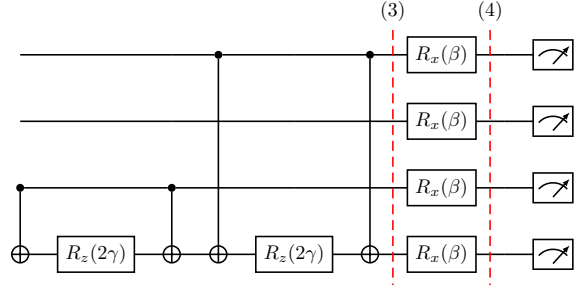

The quantum circuit diagram for the example showed in the \textbf{Figure \ref{fig:ising_spin_model}} which is the $3$ particles with linear configuration ISM problem, the main difference between this quantum circuit and the one for the $4$ nodes with cyclic configuration Max-Cut problem (besides the configuration of the graph and the number of qubits) is that this circuit has a \say{extra} section in the phase operator, this section adds the external magnetic field for the particles.

\begin{figure}[ht]
    \begin{center}
        \begin{tikzpicture}
            \node[scale=0.85] {
                \begin{quantikz}
                    \ket{0}_{1} \slice{(1)}  & \gate{H} \slice{(2)} & \ctrl{1} &\qw & \ctrl{1} & \qw & \qw & \qw \\
                    \ket{0}_{2} & \gate{H} & \targ{} & \gate{R_{z}(2\gamma)} & \targ{} & \ctrl{1} & \qw & \ctrl{1} \\
                    \ket{0}_{3} & \gate{H} & \qw & \qw & \qw & \targ{} & \gate{R_{z}(2\gamma)} & \targ{} 
                \end{quantikz}
            };
        \end{tikzpicture}
    \end{center}
\end{figure}

\begin{figure}[ht]
    \begin{center}
        \begin{tikzpicture}
            \node[scale=0.85] {
                \begin{quantikz}
                    \ldots & \qw \slice{(3)} & \gate{R_{z}(h_{1}\gamma)} & \qw \slice{(4)} & \gate{R_{x}(\beta)} & \qw \slice{(5)} & \meter{} \\
                    \ldots & \qw & \gate{R_{z}(h_{2}\gamma)} & \qw & \gate{R_{x}(\beta)} & \qw & \meter{}\\
                    \ldots & \qw & \gate{R_{z}(h_{3}\gamma)} & \qw & \gate{R_{x}(\beta)} & \qw & \meter{}
                \end{quantikz}
            };
        \end{tikzpicture}
    \caption{Quantum circuit for the $3$ particles with linear configuration ISM problem.}
    \label{fig:ISM_QuantumCircuit_3p_LIN}
    \end{center}
\end{figure}
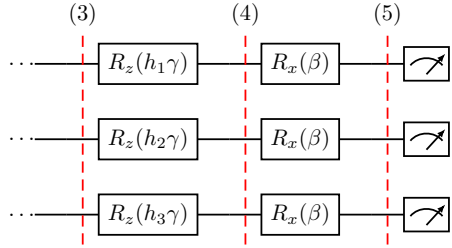

\section{Optimization methods}

In this paper, we tested two different methods to optimize the parameters for the phase operator and the mixing operator in QAOA. As it was mentioned, these two methods are classical optimization approaches that are used to generate the best combination of parameters for QAOA. \cite{12}

\subsection{Exhaustive Search}

To prove that the QAOA algorithm approach works, two kinds of simulations were developed. The first approach tested was the Exhaustive Search (ES), this method is not used to probe the benefits of this technique because is already known that all this type of brute force algorithms are good to explore a search space (through a certain degree) but all of them are really expensive in terms of time and computational resources. The idea behind the use of ES is to create a base solution which can be compared to the other approach, to know if the other optimization method generates a good solution or if the solution is completely different from the one we expect. \cite{17} \cite{12} \cite{13}

\begin{figure}[ht]
\centering
\captionsetup{justification=centering}
\includegraphics[width=8cm,height=5cm]{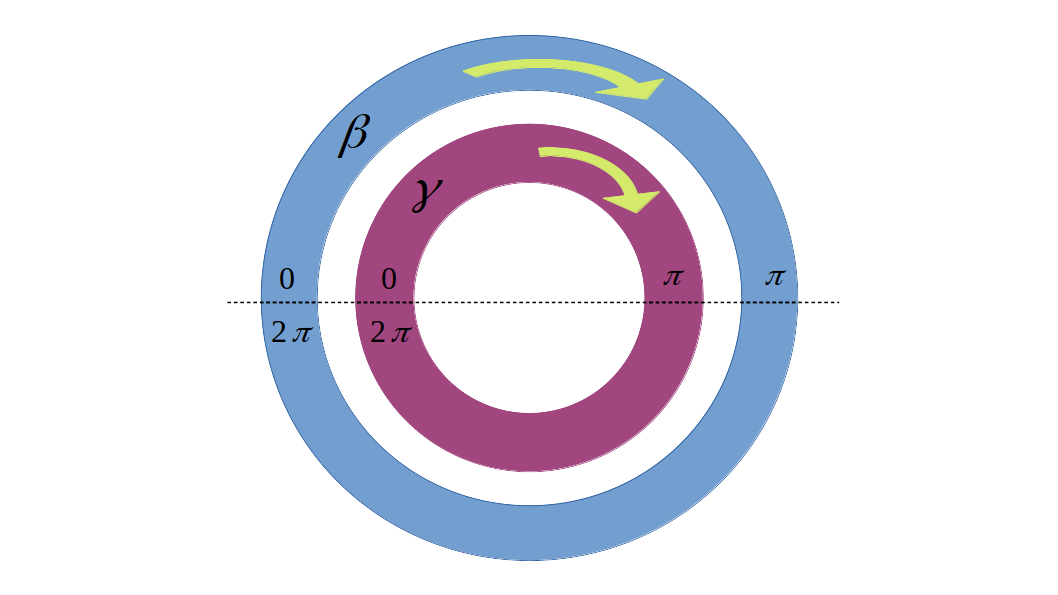}
\caption{Exhaustive Search (ES) illustration.}
\label{fig:ES_ilustration}
\end{figure}

In the \textbf{Figure \ref{fig:ES_ilustration}} is represented how the ES method works, in the base model of the QAOA there are two parameters to be optimized, each parameter can be thought as a variable that takes one value from a range of values at a time. In ES every parameter explored is going to take all the values of the range allowed, the only restriction that ES has is the precision of the values, in the sense that the range of values for the parameters in QAOA is a continued set of values from $0$ to $2 \pi$ and ES is limited to explore a discrete number of values of that set.

\subsection{Iterative Local Search }

The heuristic of Iterative Local Search (ILS) or Iterated Local Search next method to be applied and analyzed in QAOA, this heuristic is based on an improved version of the Stochastic Hill Climbing (SHC) with random starts. \cite{12} \cite{13} \cite{14} \cite{15}

\begin{figure}[ht]
\centering
\captionsetup{justification=centering}
\includegraphics[width=8cm,height=5cm]{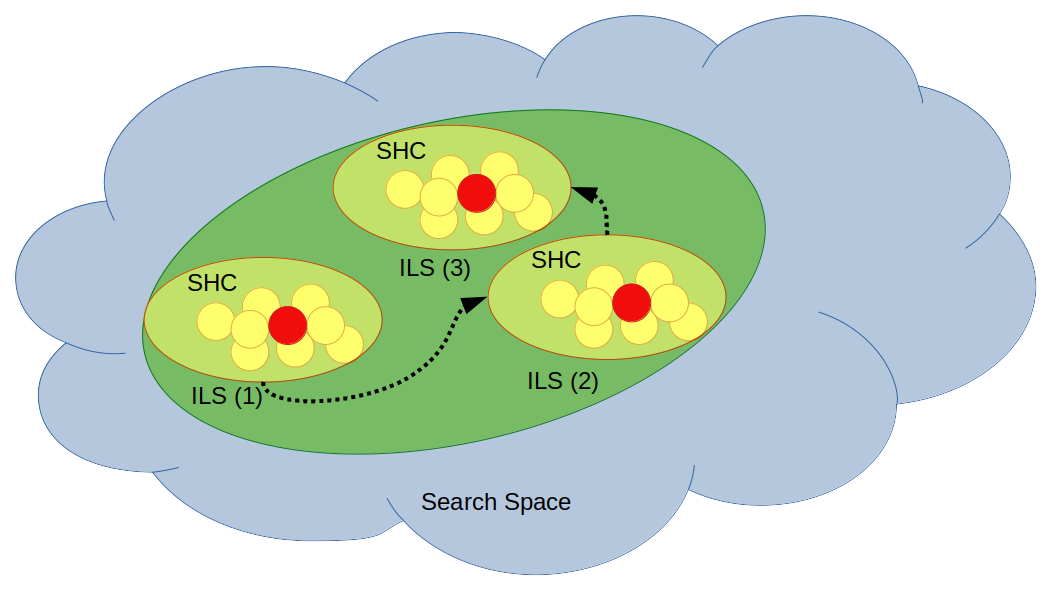}
\caption{Iterative Local Search (ILS) illustration.}
\label{fig:ILS_ilustration}
\end{figure}

The \textbf{Figure \ref{fig:ILS_ilustration}} illustrates the ILS approach. ILS starts with a random point in a search space (blue cloud), once the starting point is established and evaluated using SHC, in each iteration of the ILS a random point of an area (green oval) is selected (relatively near the starting point), this selection of random points in an area is called \say{Local Search}, using this random points then the SHC algorithm is applied to find the best point in an even more reduced area, this process continues until the number of iterations of ILS is reached. The green area that represents the local search moves (or changes its amplitude) inside the search space during the ILS iteration, in other word, the local search area is not static (doesn't have a constant area). \cite{16}

\section{Simulations and Results}

The simulation and results were separated in two subsections, the simulations using Exhaustive Search (ES) are on the first subsection and the Iterative Local Search are developed on the second subsection. Both of the classical optimization approaches were tested in local and real simulations. 

The application of the classical optimization method is reserved only for the local simulations, that is because IBM only allows a limited time for running a quantum circuit in their devices and since running the classical optimization is always going to be executed on a classical computer, the differences between running the optimization method on a classic local device and in a classic computer from IBM are practically negligible. With that said, the best combinations for the values of the QAOA parameters obtained for each problem in the local simulations are the combinations of values used on the real simulations.

Another important mention to say is related to the type of quantum computers used in the real simulations, because the majority of the quantum circuits programs were tested using the quantum computer \textbf{ibmq\_manila}, but some simulations were executed on the quantum computer \textbf{ibmq\_santiago}, these two quantum computers have different quantum processors and technologies, and these differences help also to analyze the contrast between them.

\subsection{Exhaustive Search Simulations}

Starting with the ES, all the results for all the problems in the local and real simulations were compiled in the next table.

\begin{table}
\scriptsize
\centering
\begin{tabular}{|c|r|r|r|c|c|} 
\hline
\multicolumn{6}{|c|}{{\cellcolor[rgb]{1,0.714,0.424}}\textbf{Exhaustive Search}}                                                                                                                                                                             \\ 
\hline
   & \multicolumn{5}{c|}{{\cellcolor[rgb]{0.867,0.91,0.796}}\textbf{ISM for 3 particles}}                                                                                                                                                                    \\ 
\hline
   & \multicolumn{1}{c|}{EEV Local} & \multicolumn{1}{c|}{EEV Real} & \multicolumn{1}{c|}{Optimum} & \multicolumn{1}{l|}{{\cellcolor[rgb]{0.878,0.761,0.804}}\textbf{Opt-Loc}} & \multicolumn{1}{l|}{{\cellcolor[rgb]{0.878,0.761,0.804}}\textbf{Opt-Real}}  \\ 
\hline
2p & -2.8496                        & -2.64919                      & -3.5                         & \textit{-0.6504}                                                          & \textit{-0.85081}                                                           \\ 
\hline
   & \multicolumn{5}{c|}{{\cellcolor[rgb]{0.867,0.91,0.796}}\textbf{ISM for 4 particles}}                                                                                                                                                                    \\ 
\hline
   & \multicolumn{1}{c|}{EEV Local} & \multicolumn{1}{c|}{EEV Real} & \multicolumn{1}{c|}{Optimum} & \multicolumn{1}{l|}{\textit{}}                                            & \multicolumn{1}{l|}{\textit{}}                                              \\ 
\hline
2p & -4.8044                        & -4.0834                       & -5.9                         & \textit{-1.0956}                                                          & \textit{-1.8166}                                                            \\ 
\hline
3p & -3.8814                        & -1.7432                       & -5.9                         & \textit{-2.0186}                                                          & \textit{-4.1568}                                                            \\ 
\hline
   & \multicolumn{5}{c|}{{\cellcolor[rgb]{0.867,0.91,0.796}}\textbf{ISM for 5 particles}}                                                                                                                                                                    \\ 
\hline
   & \multicolumn{1}{c|}{EEV Local} & \multicolumn{1}{c|}{EEV Real} & \multicolumn{1}{c|}{Optimum} & \multicolumn{1}{l|}{\textit{}}                                            & \multicolumn{1}{l|}{\textit{}}                                              \\ 
\hline
2p & -7.9012                        & -3.3192                       & -10.9                        & \textit{-2.9988}                                                          & \textit{-7.5808}                                                            \\ 
\hline
3p & -6.7092                        & -2.1216                       & -10.9                        & \textit{-4.1908}                                                          & \textit{-8.7784}                                                            \\ 
\hline
4p & -8.1286                        & -0.2806                       & -10.9                        & \textit{-2.7714}                                                          & \textit{-10.6194}                                                           \\ 
\hline
   & \multicolumn{5}{c|}{{\cellcolor[rgb]{1,0.847,0.808}}\textbf{Max-Cut for 3 nodes}}                                                                                                                                                                       \\ 
\hline
   & \multicolumn{1}{c|}{EEV Local} & \multicolumn{1}{c|}{EEV Real} & \multicolumn{1}{c|}{Optimum} & \multicolumn{1}{l|}{\textit{}}                                            & \multicolumn{1}{l|}{\textit{}}                                              \\ 
\hline
2p & 1.658                          & 1.5499                        & 2                            & \textit{0.342}                                                            & \textit{0.4501}                                                             \\ 
\hline
   & \multicolumn{5}{c|}{{\cellcolor[rgb]{1,0.847,0.808}}\textbf{Max-Cut for 4 nodes}}                                                                                                                                                                       \\ 
\hline
   & \multicolumn{1}{c|}{EEV Local} & \multicolumn{1}{c|}{EEV Real} & \multicolumn{1}{c|}{Optimum} & \multicolumn{1}{l|}{\textit{}}                                            & \multicolumn{1}{l|}{\textit{}}                                              \\ 
\hline
2p & 2.088                          & 1.9819                        & 4                            & \textit{1.912}                                                            & \textit{2.0181}                                                             \\ 
\hline
3p & 2.618                          & 2.312                         & 4                            & \textit{1.382}                                                            & \textit{1.688}                                                              \\ 
\hline
4p & 3.9819                         & 2.8439                        & 4                            & \textit{0.0181}                                                           & \textit{1.1561}                                                             \\ 
\hline
   & \multicolumn{5}{c|}{{\cellcolor[rgb]{1,0.847,0.808}}\textbf{Max-Cut for 5 nodes}}                                                                                                                                                                       \\ 
\hline
   & \multicolumn{1}{c|}{EEV Local} & \multicolumn{1}{c|}{EEV Real} & \multicolumn{1}{c|}{Optimum} & \multicolumn{1}{l|}{\textit{}}                                            & \multicolumn{1}{l|}{\textit{}}                                              \\ 
\hline
2p & 3.514                          & 2.9619                        & 6                            & \textit{2.486}                                                            & \textit{3.0381}                                                             \\ 
\hline
3p & 3.634                          & 2.608                         & 6                            & \textit{2.366}                                                            & \textit{3.392}                                                              \\ 
\hline
4p & 3.65                           & 2.636                         & 6                            & \textit{2.35}                                                             & \textit{3.364}                                                              \\
\hline
\end{tabular}
\caption{Exhaustive Search optimization compilation.}
\label{tab:ES_compilation}
\end{table}

Starting with the \textbf{Table \ref{tab:ES_compilation}} description, on the far left column is the type of model used for QAOA, the types can be $2p$, $3p$ or $4p$ each type stands for the number of parameters used in the operators for QAOA, for the case $2p$ it was used one parameter called $\gamma$ for the phase operator and $\beta$ for the mixing operator, for the $3p$ we also have one parameter $\gamma$ but it was used two parameters $\beta_{1}$ and $\beta_{2}$ for the mixing operator, and finally for the $4p$ it was used two phase operators and two mixing operators connected consecutively, the $4p$ case has two $\gamma$ parameters (for each phase operator) and two $\beta$ parameters (for each mixing operator).

Each of the problem's results for the ES method were compiled on the table from before, for every simulation generated it was calculated a value called EEV which stands for \textbf{Expected Energy Value}, this EEV value is very similar to the Expectation Value for the QAOA general implementation, EEV was obtained for the local and real simulations (EEV Local and EEV Real respectively). The main focus of QAOA is two approximate the EEV to the optimum value calculated, the optimum value represents that you can measure the optimal state with $100\%$ probability, if the EEV is near the optimum that tell us the probability to measure the optimal state is greater, and also the probability to measure states near the optimal is greater.

In the last two columns \textbf{Opt-Loc} and \textbf{Opt-Real} is the difference between the local and real results with respect to the optimum result.

\subsection{Iterative Local Search Simulations}

Now, for the Iterative Local Search (ILS) method, it was used the same table distribution as in the ES method.

\begin{table}
\scriptsize
\centering
\begin{tabular}{|c|r|r|r|c|c|} 
\hline
\multicolumn{6}{|c|}{{\cellcolor[rgb]{1,0.714,0.424}}\textbf{Iterative Local Search}}                                                                                                                              \\ 
\hline
   & \multicolumn{5}{c|}{{\cellcolor[rgb]{0.867,0.91,0.796}}\textbf{ISM for 3 particles}}                                                                                                                          \\ 
\hline
   & \multicolumn{1}{c|}{EEV Local} & \multicolumn{1}{c|}{EEV Real} & \multicolumn{1}{c|}{Optimum} & {\cellcolor[rgb]{0.878,0.761,0.804}}\textbf{Opt-Loc} & {\cellcolor[rgb]{0.878,0.761,0.804}}\textbf{Opt-Real}  \\ 
\hline
2p & -2.9734                        & -2.4612                       & -3.5                         & \textit{-0.5266}                                     & \textit{-1.0388}                                       \\ 
\hline
   & \multicolumn{5}{c|}{{\cellcolor[rgb]{0.867,0.91,0.796}}\textbf{ISM for 4 particles}}                                                                                                                          \\ 
\hline
   & \multicolumn{1}{c|}{EEV Local} & \multicolumn{1}{c|}{EEV Real} & \multicolumn{1}{c|}{Optimum} & \multicolumn{1}{l|}{\textit{}}                       & \multicolumn{1}{l|}{\textit{}}                         \\ 
\hline
2p & -4.8908                        & -4.1362                       & -5.9                         & \textit{-1.0092}                                     & \textit{-1.7638}                                       \\ 
\hline
3p & -4.0956                        & -3.3936                       & -5.9                         & \textit{-1.8044}                                     & \textit{-2.5064}                                       \\ 
\hline
4p & -4.7374                        & -1.99739                      & -5.9                         & \textit{-1.1626}                                     & \textit{-3.90261}                                      \\ 
\hline
   & \multicolumn{5}{c|}{{\cellcolor[rgb]{0.867,0.91,0.796}}\textbf{ISM for 5 particles}}                                                                                                                          \\ 
\hline
   & \multicolumn{1}{c|}{EEV Local} & \multicolumn{1}{c|}{EEV Real} & \multicolumn{1}{c|}{Optimum} & \multicolumn{1}{l|}{\textit{}}                       & \multicolumn{1}{l|}{\textit{}}                         \\ 
\hline
2p & -8.3734                        & 0.8364                        & -10.9                        & \textit{-2.5266}                                     & \textit{-11.7364}                                      \\ 
\hline
3p & -7.3026                        & -0.6258                       & -10.9                        & \textit{-3.5974}                                     & \textit{-10.2742}                                      \\ 
\hline
4p & -7.3473                        & 0.30479                       & -10.9                        & \textit{-3.5527}                                     & \textit{-11.20479}                                     \\ 
\hline
   & \multicolumn{5}{c|}{{\cellcolor[rgb]{1,0.847,0.808}}\textbf{Max-Cut for 3 nodes}}                                                                                                                             \\ 
\hline
   & \multicolumn{1}{c|}{EEV Local} & \multicolumn{1}{c|}{EEV Real} & \multicolumn{1}{c|}{Optimum} & \multicolumn{1}{l|}{\textit{}}                       & \multicolumn{1}{l|}{\textit{}}                         \\ 
\hline
2p & 1.716                          & 1.5049                        & 2                            & \textit{0.284}                                       & \textit{0.4951}                                        \\ 
\hline
   & \multicolumn{5}{c|}{{\cellcolor[rgb]{1,0.847,0.808}}\textbf{Max-Cut for 4 nodes}}                                                                                                                             \\ 
\hline
   & \multicolumn{1}{c|}{EEV Local} & \multicolumn{1}{c|}{EEV Real} & \multicolumn{1}{c|}{Optimum} & \multicolumn{1}{l|}{\textit{}}                       & \multicolumn{1}{l|}{\textit{}}                         \\ 
\hline
2p & 2.144                          & 1.956                         & 4                            & \textit{1.856}                                       & \textit{2.044}                                         \\ 
\hline
3p & 2.6919                         & 2.156                         & 4                            & \textit{1.3081}                                      & \textit{1.844}                                         \\ 
\hline
4p & 4                              & 2.498                         & 4                            & \textit{0}                                           & \textit{1.502}                                         \\ 
\hline
   & \multicolumn{5}{c|}{{\cellcolor[rgb]{1,0.847,0.808}}\textbf{Max-Cut for 5 nodes}}                                                                                                                             \\ 
\hline
   & \multicolumn{1}{c|}{EEV Local} & \multicolumn{1}{c|}{EEV Real} & \multicolumn{1}{c|}{Optimum} & \multicolumn{1}{l|}{\textit{}}                       & \multicolumn{1}{l|}{\textit{}}                         \\ 
\hline
2p & 3.6879                         & 2.6859                        & 6                            & \textit{2.3121}                                      & \textit{3.3141}                                        \\ 
\hline
3p & 3.7499                         & 2.484                         & 6                            & \textit{2.2501}                                      & \textit{3.516}                                         \\ 
\hline
4p & 3.758                          & 2.828                         & 6                            & \textit{2.242}                                       & \textit{3.172}                                         \\
\hline
\end{tabular}
\caption{Iterative Local Search optimization compilation.}
\label{tab:ILS_compilation}
\end{table}

The \textbf{Table \ref{tab:ILS_compilation}} has all the values from the simulations of the problems for ISM and Max-Cut using ILS in local and real experiments, the only difference with ES table is that for the ISM $4$ particles problem it was also simulated using $4p$ model of QAOA.

\subsection{Comparative average results for the classical optimization methods}

Using the results of the \textbf{Tables \ref{tab:ES_compilation}} and \textbf{\ref{tab:ILS_compilation}} two comparative tables were made. The first table has the average values for the Exhaustive Search optimization method.

\begin{table}
\centering
\begin{tabular}{|c|c|} 
\hline
\multicolumn{2}{|c|}{{\cellcolor[rgb]{1,0.714,0.424}}\textbf{Exhaustive Search}}  \\ 
\hline
\textit{Average Opt-Loc}              & \textit{Average Opt-Real}                 \\ 
\hline
\rowcolor[rgb]{0.867,0.91,0.796} ISM  & ISM                                       \\ 
\hline
\textit{\textbf{-2.2876}}             & \textit{\textbf{-5.63}}                   \\ 
\hline
\rowcolor[rgb]{1,0.847,0.808} Max-Cut & Max-Cut                                   \\ 
\hline
\textit{\textbf{1.55}}                & \textit{\textbf{2.16}}                    \\
\hline
\end{tabular}
\caption{Average comparative results for ES.}
\label{tab:ES_comparative}
\end{table}

Now, the next table shows the average results for the ILS method.

\begin{table}
\centering
\begin{tabular}{|c|c|} 
\hline
\multicolumn{2}{|c|}{{\cellcolor[rgb]{1,0.714,0.424}}\textbf{Iterative Local Search}}  \\ 
\hline
\textit{Average Opt-Loc}              & \textit{Average Opt-Real}                      \\ 
\hline
\rowcolor[rgb]{0.867,0.91,0.796} ISM  & ISM                                            \\ 
\hline
\textit{\textbf{-2.03}}               & \textit{\textbf{-6.06}}                        \\ 
\hline
\rowcolor[rgb]{1,0.847,0.808} Max-Cut & Max-Cut                                        \\ 
\hline
\textit{\textbf{1.46}}                & \textit{\textbf{2.27}}                         \\
\hline
\end{tabular}
\caption{Average comparative results for ILS.}
\label{tab:ILS_comparative}
\end{table}

Analyzing both \textbf{Tables \ref{tab:ES_comparative}} and \textbf{\ref{tab:ILS_comparative}} it can be seen that the ILS heuristic presents better average results for the local simulations for both ISM and Max-Cut class of problems. It is curious to see that the results for the real simulations are better with ES approach, but due to the different factors that can alter the real simulations, in this work we are going to keep the ILS as a better method to be applied with QAOA. And if the complexity analysis of both classical optimization methods is taken into account, ILS still has the lead over the ES method. In all the simulations developed (local and real) the values generated for the parameters (for $2p$, $3p$ and $4p$ models) were better with ILS but when some values from ILS are implemented in the quantum computer, due to their precision (how specific is the value of that parameter), it makes harder for the quantum computer to replicate the exact quantum gate using the specific value of rotation, and that could be one of the reasons why the ES values perform better (from the experimental results) in the real simulations.

With respect to the \textbf{Tables \ref{tab:ES_compilation}} and \textbf{\ref{tab:ILS_compilation}} from the compilation results, another interesting result can be discussed, this results is shown in relation between the real and local simulations using the different QAOA models ($2p$, $3p$ or $4p$) because in the case of the local simulations with the increase of the parameters in the QAOA model most of the time the response of the algorithm was improved (the value of EEV got closer to the optimum) however for the real simulations the results usually got worst with more parameters, and these phenomena can also be seen in the \textbf{Tables \ref{tab:ES_comparative}} and \textbf{\ref{tab:ILS_comparative}} where all the average values for both ES and ILS were worst in the case of the real simulations. These two results show the state of NISQ computers compared to the ideal results (from local simulations), the actual NISQ devices still have a lot of problems with noise and decoherence, and when the number of quantum gates or qubits increase, the noise and decoherence also increases, and even when in real simulations it was used the best values for the parameters of QAOA, the local results were much better than the real simulations results.

The final result obtained is from the compilation tables for ES and ILS, because the majority of the results from the Max-Cut problems were closer (even the real simulation results) to the optimum. With this in mind, it can be said that the Max-Cut problems perform better with QAOA, one important factor in contrast from Max-Cut to ISM is that Max-Cut has at least two optimal states (and since QAOA is an approximation algorithm) it makes it \say{easier} for QAOA to find one or more solutions in all the search space that are approximated to the optimum, instead of ISM where QAOA needs to find only one solution. 

\section{Conclusion}

The first conclusion established from all the simulations developed in this paper is related to the classical optimization method, Iterative Local Search showed that is a better optimization heuristic to be implemented with QAOA for the ISM and Max-Cut problems. However, for the quantum circuits in the real simulations it is harder to implement the exact values obtained from ILS in the quantum gates, and in the actual state of the quantum computers, it can alter the parameter values and consequently the results may vary from the ideal expected results.

Another important conclusion, is in relation to the state of NISQ computers, these devices still have noise and decoherence problems, the experiments exhibited the difference between the local simulations that use ideal conditions to calculate the quantum circuits phenomena and the real simulations which are affected by the imprecision of today's devices. The greatest number of errors among the results generated were when there were models with more number of quantum gates and higher number of qubits.

The last conclusion, represents the difference between the types of problems that work better with QAOA. The results for Max-Cut problems were better than ISM problems, it can be said that is easier for QAOA to approximate the EEV to the optimum for those problems because QAOA is an approximation optimization algorithm and allowing QAOA to be able to find more than one solution in a search space is simpler than finding only one optimum (like in ISM). Also, as it was mentioned, the Max-Cut problem translation to a quantum circuit needs less quantum gates and this allows, in the case of the real simulations, to maintain a lower number of quantum gates compared with the ISM problems which translates into a response that is closer to the optimum.

\section*{Acknowledgment}

Thanks to Universidad Aut\'onoma Metropolitana (Unidad Azcapotzalco), for the education given for the development of this article. 

Thanks to Fis. Jos\'e de Jes\'us Cruz Guzm\'an and Professor Zbigniew Oziewicz for the initiation of the quantum computing paradigm and all the priceless talks in different areas of physics and mathematics.

Thanks to Dr. Alvaro Salas Brito for the mentorship on my master's degree thesis, and also for teaching the quantum computing fundamentals, which helped in the development of this paper.

And, thanks Dr. Roman Anselmo Mora Gutierrez for the guidance on my master's degree, helping to understand the basics of heuristics and combinatorial optimization problems.

\printbibliography

\end{document}